A.A.Kozlov

# ON PHYSICAL NATURE OF RADIATION REGULATING QUANTITY OF CELLS IN POPULATION

*Laboratory of Development Biology, Department of Biology, Tbilisi State University, 01043, Tbilisi, Georgia*
E-mail: <alekskoz@rambler.ru>

Our previous experiments have shown that a certain physical radiation restricts the quantity of cells (infusoria) in developing culture "from above". Given report describes the series of experiments which enabled to determine the nature of this phenomenon. Occurred that this radiation belongs to UV spectral region with the wave range (200-290) nm.

Defined has been influence of concentration of nutrient medium on dynamics of development of unicellular culture conditioned by different transparency of different concentration nutrient medium to the UV radiation.

As reported in [1,2], total quantity of cells (infusoria) in the culture is regulated by radiation originating from these cells, however, the nature if this radiation remains unclear.

In principle the mentioned regulating radiation may have either electromagnetic or sound nature. At that, this radiation is to be outside the spectrum (electromagnetic and sound) which naturally exists in environment; otherwise, the signal/noise ratio will be low for reliable isolation of desired signal. Besides, radiation energy must be enough for implementing the regulatory function. Therefore, we assumed that sound radiation of the cells and their electromagnetic radiation in far infrared area are not likely to trigger observed effect.

## Materials and methods

For the first series of experiments, in order to define the region of electromagnetic spectrum able to exert regulatory function in the cells of the culture the vessels made of plexiglas divided into three sections have been used. Compartment A (volume 5 $cm^3$) was separated from compartment B (volume 150 $cm^3$) with 5 mkm thick fluoroplastic film, whereas compartment C (volume 5 $cm^3$) and compartment B were separated with 1 mm thick plexiglas plate – see Chamber 1 (Figure 1a). In the second series the plates were replaced with 1 mm thick quartz cuvette A with two matted opposite facets and 1 mm thick plexiglas cuvette C. Operational volumes of the cuvettes are 5 $cm^3$, volume of compartment B – 150 $cm^3$ – see Chamber 2 (Figure 1b).

Tightness of fluoroplastic film has been checked using luminescent dye thripaphlavine solution. The transmission spectrum of the fluoroplastic film, quartz and plexiglas are given in Figure 2.

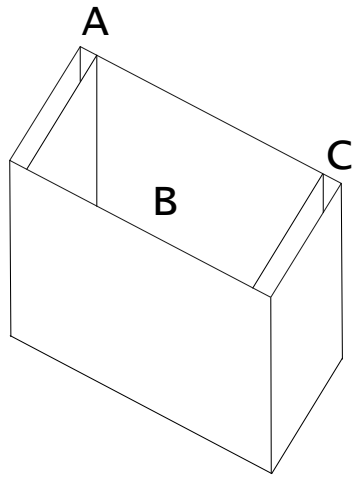 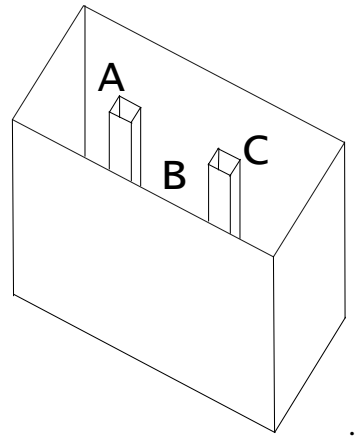

**Figure 1a.**  **Figure 1b.**
**Chamber 1.**  **Chamber 2.**

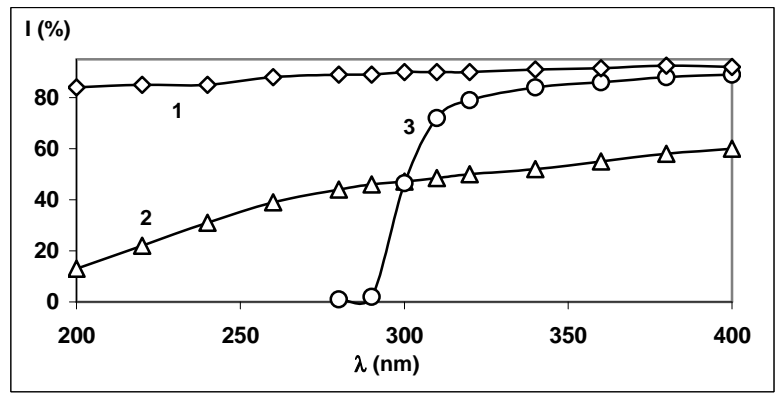

**Figure 2.  Transmission spectrum of quartz, fluoroplast and plexiglas**
**curve 1- quartz, curve 2 - fluoroplast, curve 3 - plexiglas**

As given in Figure 2, one millimeter thick plexiglas cuts off the spectral region with λ ≤ 290 nm. In this area transmission of the 5 mkm thick fluoroplast film changes from 47% to 13%, for the quartz cuvette reduces from 90% to 84%.

In all of the three compartments (or cuvettes) nutrient medium (hay broth) with infusoria *Colpoda* or *Paramecia* with similar initial concentration of cells at the stage of exponential growth has been poured in. In 12-16 days, when the culture enters into the quasi-permanent phase of development, concentration of the cells in the mentioned compartments has been measured.

### Results and discussion

In all experiments small concentration of cells in quasi-stationary phase has been observed in compartment C, as it has been expected to be in the small volumes [1,2].

Average concentration of infusoria *Colpoda* in compartment C in this phase is in the order of $10^3$ cm$^{-3}$, whereas for infusoria *Paramecia* the value is about 50 cm$^{-3}$. At the same time it should be mentioned that these values vary from test to test. Therefore, for the clearness of quantitative results in each case, irrespective to the species of infusoria, concentration of cells in compartment C was recognized as 1, while concentrations in other compartments has been expressed as fraction of 1. Figure 3 shows results of the series of experiments.

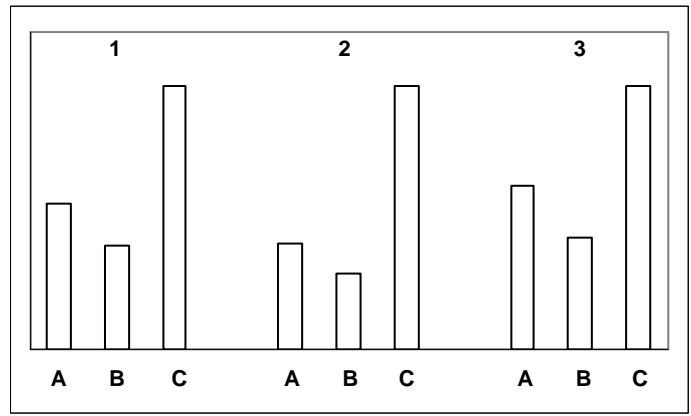

**Figure 3. Relative concentration of cells in quasi-stationary phase of development according to the series of experiments.**
**Vessel 1 –(Figure 1a), *Colpoda*, 5 mkm fluoroplastic film partition (5 experiments).**
**Vessel 2 - (Figure 1a), *Paramecia*, 5 mkm fluoroplastic film partition (2 experiments).**
**Vessel 3 - (Figure 1б), *Paramecia*, 1 mm quartz cuvette (3 experiments).**
**A, B and C – concentration of cells in the mentioned compartments and cuvettes.**

As given in the figure above, qualitatively the situation in all of the cases is identical, as for the qualitative difference - it is rather slight, so all the results can be combined into one histogram which as given in Figure 4.

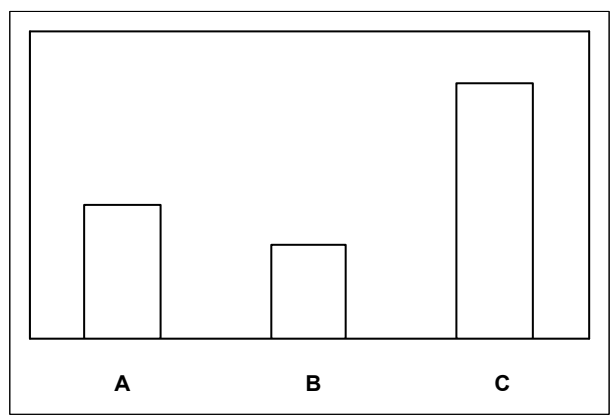

**Figure 4. Relative concentrations of cells in quasi-stationary phase in volumes A, B and C averaged by all series**

In the article [1,2] assumption is made that the factor regulating the upper quantity of the cells in the culture is a certain radiation generated by the cells themselves when a volumetric density reaches a critical value after which no subsequent cell division happens. Average volumetric density of this radiation is directly proportional to the volume where the cells are cultivated and inversely proportional to the surface area through which the radiation is lost. Therefore, critical density of radiation "switching off" the cell division in smaller volumes is attained at higher concentration of cells than in larges volumes.

Figure 4 shows that the lowest concentration is observed in compartment B as it has been expected for the larger volumes. In the similar size compartments and cuvettes A and C concentration of cells is different: average concentration of cells in compartment C versus that in compartment A makes $C_C/C_A = 1.95 \pm 0.17$. Ultraviolet constituent does not get from C to B compartment since it can not penetrate the plexiglas barrier. Therefore, concentration of cells in this compartment is similar to that which should be observed in smaller compartments ($C_C/C_B = 2.80 \pm 0.32$). At the same time, a certain portion of UV constituent gets from compartment B into A where it sums up with cells radiationю Because of that, critical density of the flow of regulating radiation in this volume is achieved at lower concentration of the cells: $C_A/C_B = 1.50 \pm 0.16$.

The following control has been performed: similar concentration of *Paramecia* culture were placed in compartments A and C, compartment B was filled up with pure nutrient medium. The test was performed three times and revealed that average ratio of cell's concentration makes: $C_A/C_C = 1.06 \pm 0.09$. This means that the difference in concentration of cells in A and C compartments, when B is filled up with infusoria culture, is due to the penetration of UV constituent of the cells radiation through the fluiriplastic membrane or the quartz cuvette.

Thus, we can assert that radiation, regulating quantity of cells in infusoria culture, is of electromagnetic nature and belongs to the UV spectral region between 290 and 200 nm. Wave lengths below 200nm are strongly absorbed by nutrition medium, so their participation in the process is less likely.

After the mentioned results were obtained it was assumed advisable to check how the nutrition medium of different optical density changes dynamics of development of the culture. In order to examine that nutrition mediums with two different concentrations were selected. As the first medium a broth - 5 g of dry *Melilotus officinalis* per one liter of water (this concentration has been assumed as 100%) had been prepared: The second nutrient medium was made by four fold dilution of this broth (25% solution). The measured transmission through 10mm medium stratum is shown in Figure 5. The chart presents transmission in per cents versus wave length in nm.

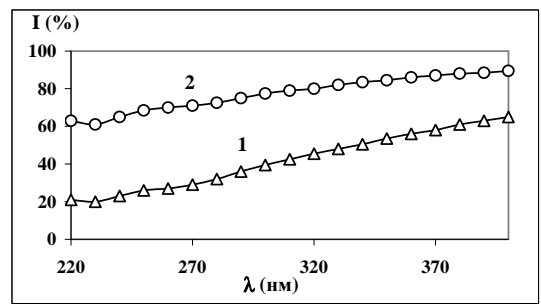

**Figure 5. Transmission spectra for nutrition media of different concentration
Curve 1 – concentrated medium (100%), curve 2 – diluted nutrition medium (25%).**

The figure shows that even in the short wave part of presented spectra nutrient medium is rather transparent. On these two media placed in cylindrical vessels by 50 ml in each (three vessels per one medium) infusoria *Colpoda* has been cultivated. Average concentration versus time is plotted in Figure 6.

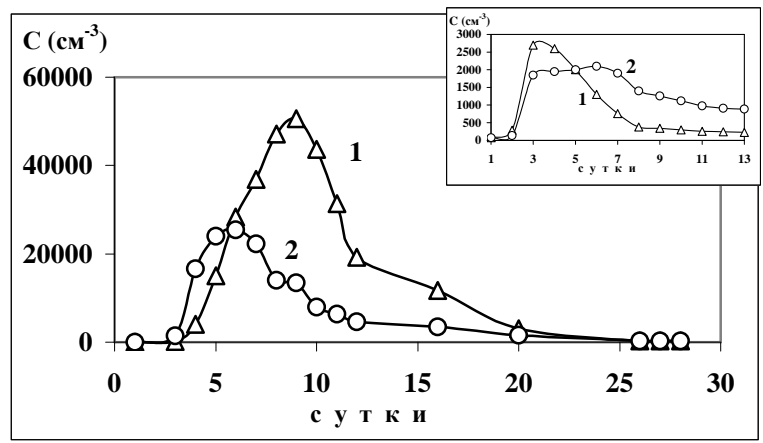

**Figure 6. Average concentration of cells versus time for different nutrition medium Curve 1 – concentrated medium, curve 2 - diluted. For comparison, in the right upper corner embedded is concentration of cells versus time in two different capacity vessels [1]. Curve 1 corresponds to 30 cm$^3$, curve 2 – for 3 cm$^3$ chambers**.

After these results were obtained we have got a possibility to make some conclusions.

As mentioned in [1,2], at initial stage of development of the cultures in different size (volume) vessels that in the larger volume developes faster. This phenomenon has been explained by induced triggering of cell division since the density of inducing radiation in larger vessels is higher than in the smaller ones. Figure 6 shows, that at the early phase of development of the culture in identical size chambers development in the chamber with lower concentration of nutrition medium goes faster. The effect can be explained by the fact, that UV photons, triggering cells division in concentrated nutrition medium absorb at smaller distances.

Besides, the figure shows, that in the chambers with concentrated medium maximum concentration of the cells is higher than that observed in the chambers with diluted broth whereas if the concentration of medium is identical concentration of cells occurs to be higher in the larger vessels (see embedded chart, Figure 6). The reason for that is the same: photons regulating quantity of cells in the culture belong to the UV spectral region, thus, in order to reach this density of irradiation required for realization of this function, concentration of cells in larger volumes and in concentrated medium must be higher than in the lower concentration medium. Conclusion can be made that photons triggering cells division (wave length 250nm) and photons, regulating quantity of cells in the culture have energy of the same order of magnitude (or are even similar)..

**Conclusion**

Thus, "dark or superweak chemiluminescence" of cells the study of which has been initiated by A.G.Gurvich and his school and continued in the 60'ies (for example see [3]) has no effect on staring of cells division process [4,5] but is the factor restricting maximum quantity of cells, i.e. radiation is not merely a byproduct of cells metabolism, but has significant biological function.